\documentclass[aps,prl,twocolumn, amsmath,amssymb,footinbib,superscriptaddress]{revtex4-1}

\usepackage{graphicx}
\usepackage{gensymb}
\usepackage{dcolumn}
\usepackage{bm}
\usepackage{color}
\graphicspath{ {./figs/} }

\begin{document}

\preprint{APS/123-QED}

\title{Bandstructure and Contact Resistance of Carbon Nanotubes Deformed by the Metal Contact}

\author{Roohollah Hafizi}
\affiliation{Skolkovo Institute of Science and Technology, 3 Nobel Street, Skolkovo, Moscow Region 143025, Russia}
 \affiliation{Department of Physics, Isfahan University of Technology, Isfahan 84156-83111, Iran.}
\author{Jerry Tersoff}%
\affiliation{%
 IBM T.J. Watson Research Center, Yorktown Heights, New York 10598, USA}
\author{Vasili Perebeinos}
\email[]{v.perebeinos@skoltech.ru}
\affiliation{Skolkovo Institute of Science and Technology, 3 Nobel Street, Skolkovo, Moscow Region 143025, Russia}
%




\date{\today}

\begin{abstract}
Capillary and van der Waals forces cause nanotubes to deform or even collapse under metal contacts.
Using ab-initio bandstructure calculations, we find that these deformations reduce the bandgap by as much as 30\%, while fully collapsed nanotubes become metallic. Moreover degeneracy lifting, due to the  broken axial symmetry and wavefunctions mismatch between the fully collapsed and the round portions of a CNT, leads to a three times higher contact resistance. The latter we demonstrate by contact resistance calculations within the tight-binding approach.

\begin{description}

\item[PACS numbers]

\end{description}
\end{abstract}

\maketitle


The fascinating mechanical and electrical properties of carbon nanotubes (CNTs) have attracted a lot of attention for a variety of technologies \cite{yakobson1996nanomechanics, falvo1997bending, smalley2003carbon, franklin2012sub, de2013carbon, geier2015solution, cao2015end, Qiu2017_Science_CNT}.
In particular, semiconducting single-wall CNT field-effect transistors have been considered for sub 10 nm technology
nodes~\cite{cao2015end, franklin2012sub, Qiu2017_Science_CNT, Cao_Science_2017}.
CNTs provide high-performance channels
below 10 nanometers, but the increase in contact resistance with decreasing
size dominates the performance of scaled devices as the channel transport becomes ballistic.
Only recently low-resistance end-bonded contacts have been demonstrated \cite{cao2015end,Cao_Science_2017}, while most commonly used Pd contacts establish side contacts~\cite{Franklin_NatNano_2010}.
In the latter, an electron has to overcome two barriers: between the metal and the nanotube under the metal and between the nanotube under the metal and the nanotube in the channel.
While much efforts have been made to  describe the former type of barriers~\cite{Leonard_NatNano_2011}, less attention has been drawn to the latter.

Although CNTs have a very large Young's modulus in the axial direction \cite{yu2000strength},
they are rather soft in the radial direction \cite{palaci2005radial, yu2000investigation}, such that they can be deformed by the influence of van der Waals forces on two adjacent CNTs \cite{ruoff1993radial}. These deformations are predicted to be much stronger in CNTs partially covered by a metal contact, due to capillary forces \cite{perebeinos2014carbon}.
While much efforts in exploring electronic structure in  deformed nanotubes have been made~\cite{lammert2000_cnt_collapsed_dft,giusca2007_cnt_collapsed_stm,NishidatePRB2008,nakanishi2015_cnt_collapsed_effectivemass} including effects of the external transverse electric field~\cite{GunlyckeEuro2006}, little is known about how deformations produced by a metal modify the electronic structure and, as a result, the contact resistance.

In this work,  we investigate the effect of such deformations on the electronic structure and contact resistance between the metal deformed and the round portions of a nanotube. Geometry relaxations are done using the valence force model~\cite{perebeinos2009valence,perebeinos2014carbon,perebeinos2015wetting}, electronic structure calculations using Density Functional Theory (DFT)~\cite{wien2K}, and contact resistance calculations within the tight-binding approach~\cite{groth2014kwant}.  In deformed, but still open semiconducting CNTs, we find bandgap reduction by 10\%-30\% depending on a tube diameter. The fully collapsed CNTs are found to be metallic. Most importantly, a new mechanism for degeneracy lifting,  arising from the interaction among the $\pi$ orbitals on adjacent sidewalls of a collapsed nanotube, leads to qualitatively larger magnitudes of the bands splitting in the fully collapsed CNTs. The band splitting is now several hundreds of meV, such that there is a dramatic impact on transport even at room temperature, unlike in previous work.

\begin{figure}
\includegraphics[scale=1]{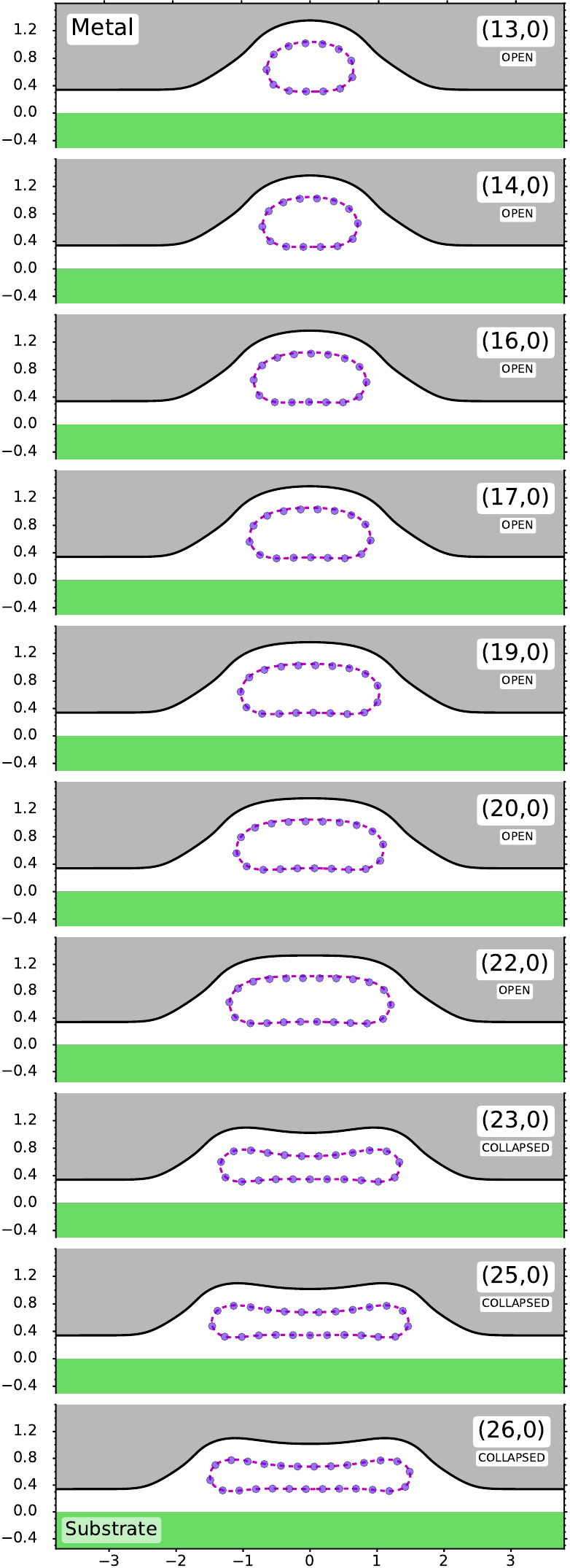}
\caption{Relaxed atomic positions (purple dots) over curvature obtained using a continuum model~\cite{perebeinos2014carbon,perebeinos2015wetting} (pink dash). Grey regions show metal contacts and black lines show their surfaces. Axes show scales in nanometers.}
\label{fig:deformed-vs-continuum}
\end{figure}

To find atomic positions of carbon atoms in CNTs under metal,
we employ a semi-atomistic model in which positions of carbon atoms are relaxed in the presence of the metal and substrate, where both are treated by a continuum model, following Ref.~\cite{perebeinos2014carbon}. The CNT interatomic interactions are described by the valence force model \cite{perebeinos2009valence},
in which stiffness against the misalignment of neighboring $\pi$ orbitals is adjusted~\cite{footnotephononmodel_adjust} to reproduce the bending stiffness of $D = 1.4~eV$ \cite{perebeinos2014carbon, Chopra1995, tomanek1993stability, blase1994europhys}.
The substrate is planar and rigid, and the metal is isotropic.
We choose surface energy $\gamma = 12.5~eV/nm^2$ as for Pd metal~\cite{singh2009surface, tyson1977surface}, the most-used contact metal.

Atomic positions are found by minimizing the total energy of the system, described by
\begin{multline}\label{energy_tot}
E = E_{bending}
+ \sum_{i,j} U_{cc}(i,j)
+ \sum_{i}\int U_{sc}(i)dS_s \\
+ \sum_{i}\int U_{mc}(i)dS_m
+ \gamma \int dS_m
+ \iint U_{ms}dS_m dS_s
\end{multline}
where \textit{i} and \textit{j} run over carbon atoms in CNT, and
$U_{cc}$, $U_{mc}$, and $U_{sc}$ are the van der Waals interactions describing CNT self-interaction, the metal-CNT interaction and the substrate-CNT interaction, respectively. These interactions are modelled by the usual 6-12 potential with parameters chosen to reproduce the binding energy $2.29~eV/nm^2$ (60~meV/carbon atom) and equilibrium spacing 0.34 nm for a flat graphene sheet \cite{low2012deformation, zacharia2004interlayer, spanu2009nature}. The
fifth term is the metal surface energy and the last term is the interaction between the metal and the substrate. For simplicity, we have chosen the same parameters for all van der Waals interactions, such that the wetting angle of the metal on the substrate is 145$\degree$.

\begin{figure}
\includegraphics[scale=0.4]{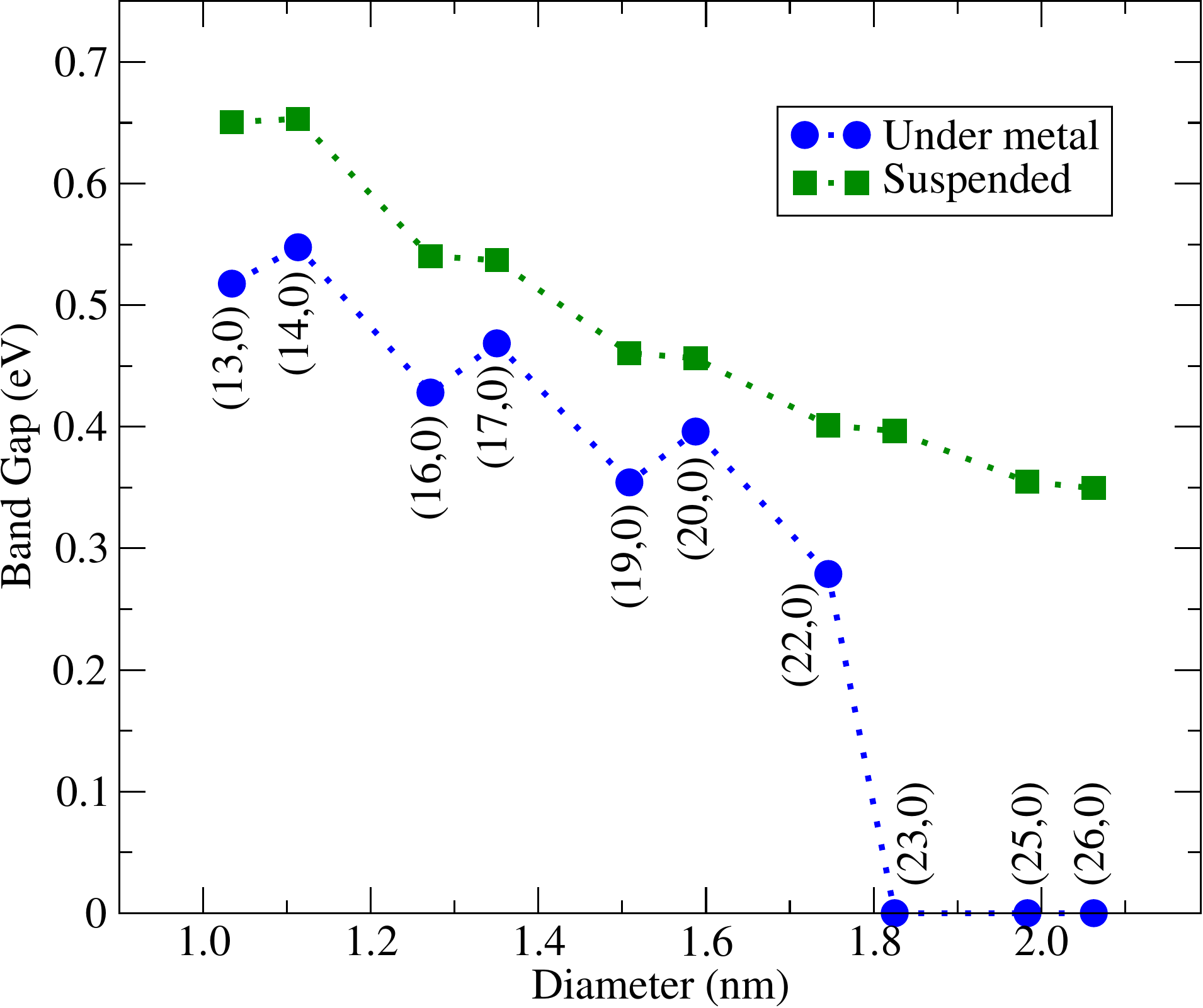}
\caption{Bandgap values from DFT of the round and metal-deformed zig-zag CNTs as a function of CNT diameter.}
\label{fig:gap-deformed-vs-round}
\end{figure}

\begin{figure}
\includegraphics[scale=0.35]{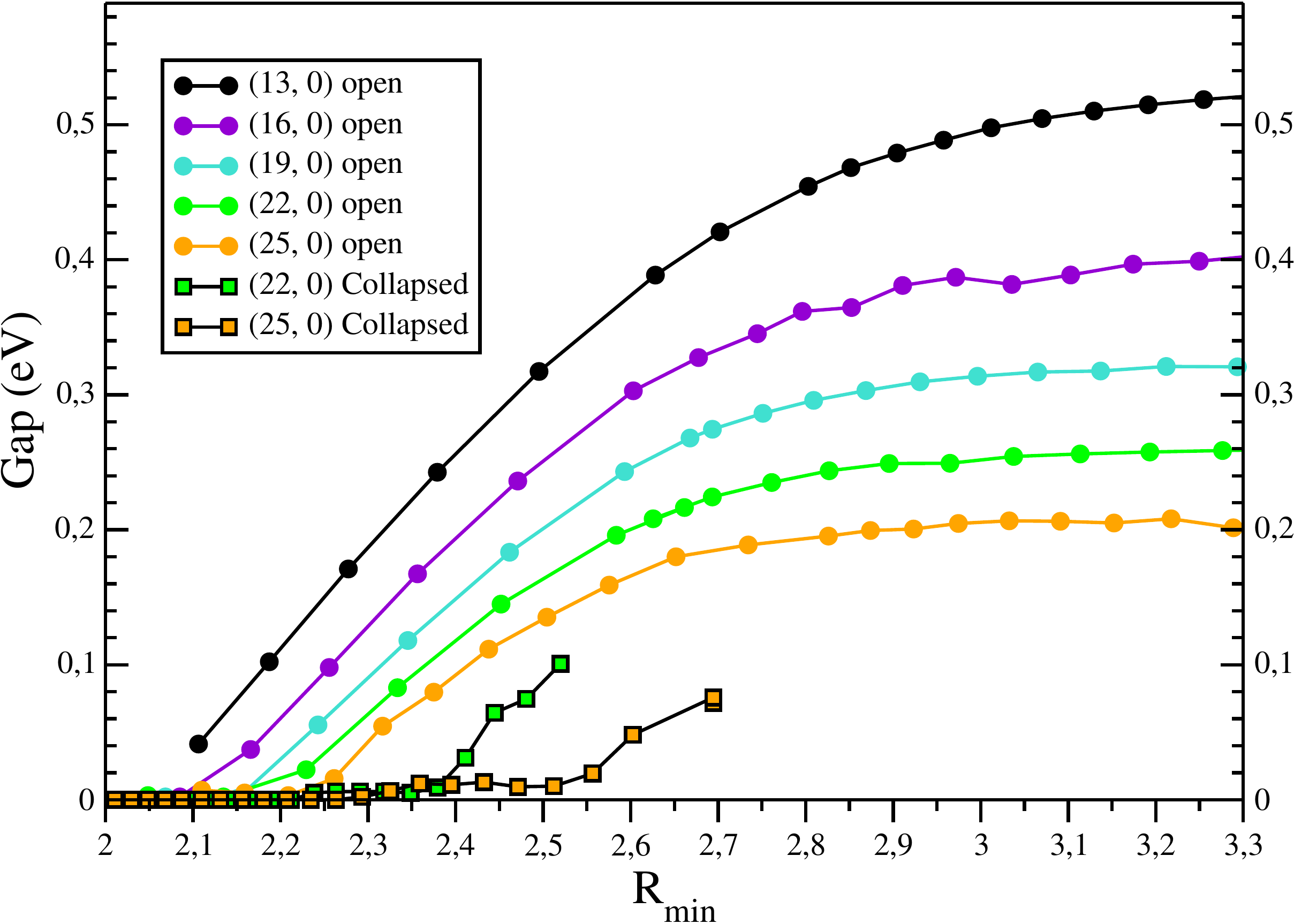}
\caption{ Bandgap values as a function of the minimum radius of curvature from DFT in the open deformed and collapsed zig-zag CNTs.}
\label{fig:gap-rmin_open_and_close}
\end{figure}

CNT geometry corresponding to the energy minimum in Eq.~(\ref{energy_tot}) is used for the electronic structure DFT calculations, employing the PBE/GGA approximation for the exchange-correlation energy~\cite{wien2K}. The vacuum used to separate CNTs is 7 \AA \ \ and the k-mesh is $1\times1\times32$.

\begin{figure}
\includegraphics[scale=0.45]{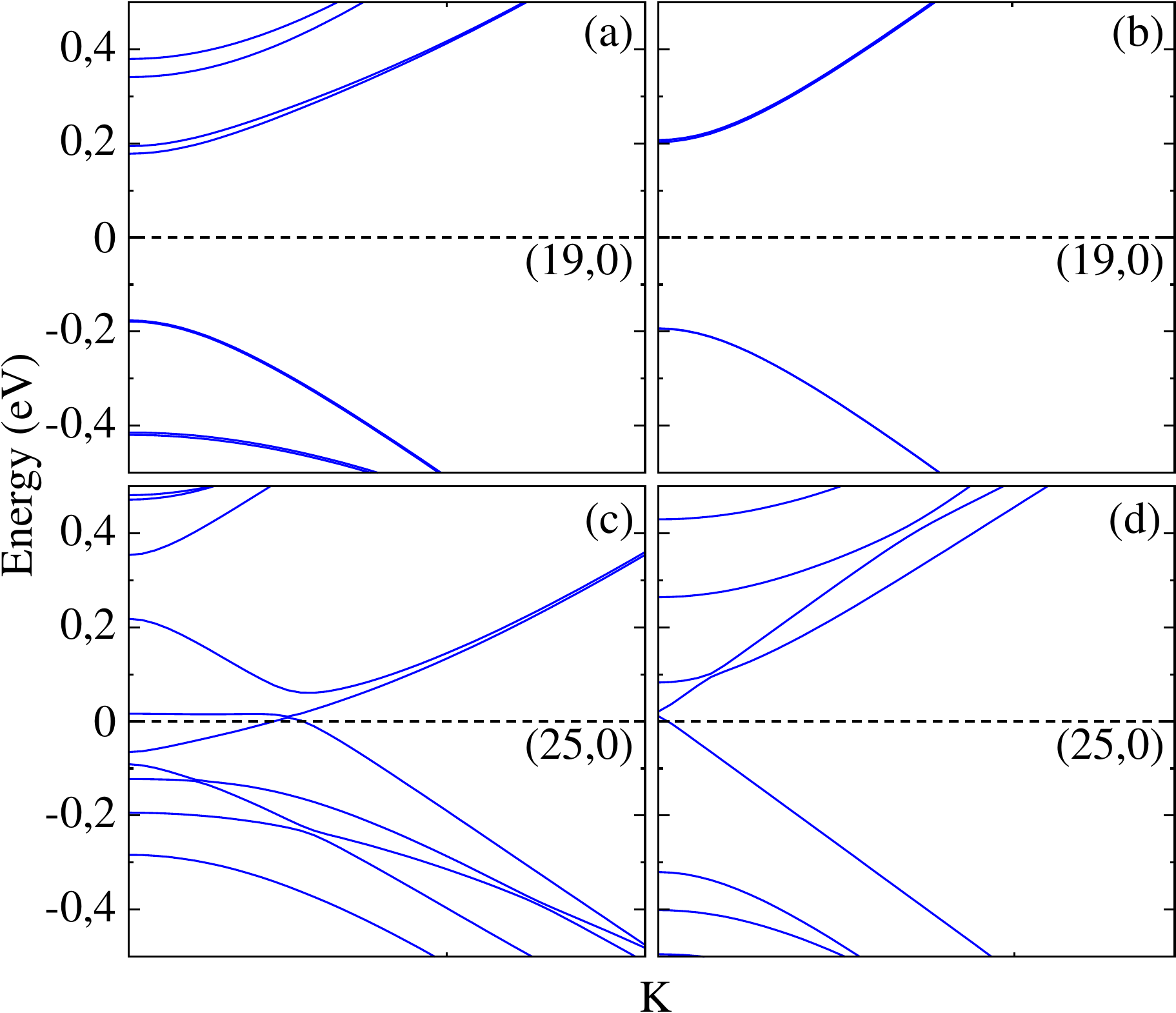}
\caption{DFT band structures of the metal-deformed (19,0) and collapsed (25,0) nanotubes in (a) and (c), correspondingly.  Tight-binding model calculations~\protect{\cite{blase1994hybridization,perebeinos2012phonon}} for the same geometries are shown in (b) and (d).
Zero of energy corresponds to the Fermi level position. Note that the double degeneracy is lifted due to the broken axial symmetry in both cases, but by a much greater amount in the fully collapsed geometry.}
\label{fig:dft-and-TB}
\end{figure}

\begin{figure}
\includegraphics[scale=0.35]{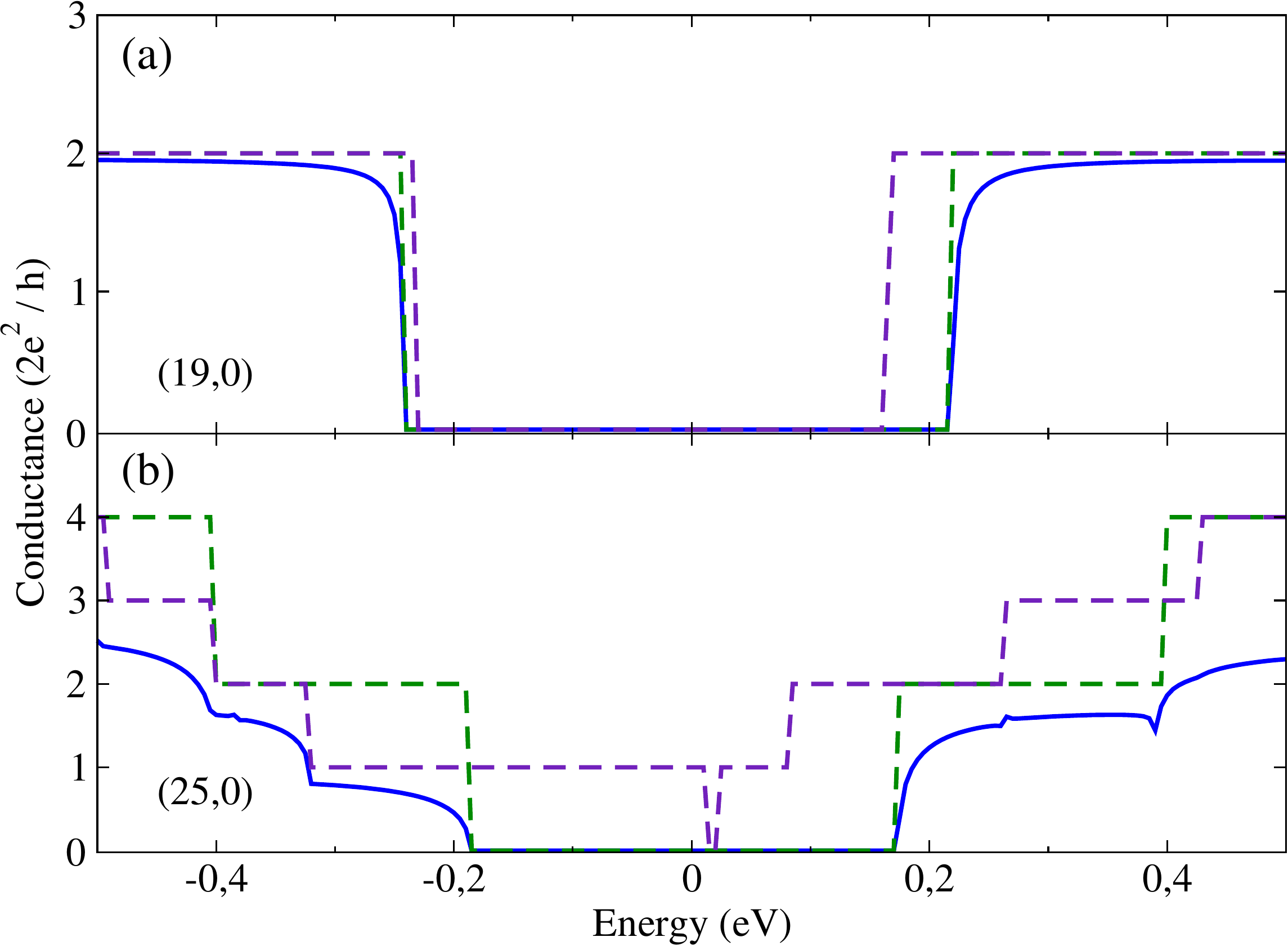}
\caption{Contact conductances as a function of energy between the suspended and metal-deformed CNTs (solid blue line), suspended-suspended CNTs (dashed green line), and deformed-deformed CNTs (dashed purple line) for (19,0) CNT in (a) and (25,0) CNT in (b).}
\label{fig:conductance}
\end{figure}

Relaxed geometries of open and collapsed zig-zag CNTs along with the results of a continuum model~\citep{perebeinos2014carbon} are shown in Fig.~\ref{fig:deformed-vs-continuum}.
DFT forces normal to the CNT circumference, i.e. along the pi-orbital directions,  correspond to a circumference-averaged pressure of about 1-2 GPa, which is consistent with the valence force model calculations in the absence of the metal and the substrate~\cite{forceissue}.
The largest open structure CNT we could find is (22,0), although metastable collapsed geometries for (19,0), (20,0), and (22,0) CNTs are found, which is consistent with Ref.~\cite{perebeinos2014carbon,perebeinos2015wetting}.
We find that atomistic and continuum model geometries in Fig.~\ref{fig:deformed-vs-continuum} are very similar. Nevertheless, in collapsed CNTs, carbon atoms in the top layer of the collapsed region prefer not to lie on top of carbons in the bottom layer, as shown in Fig.~\ref{fig:deformed-vs-continuum}. It is energetically more favorable for carbon atoms to slide by half the C-C bondlength in the curvature direction.

Fig.~\ref{fig:gap-deformed-vs-round} shows DFT bandgaps in suspended, i.e. round CNTs relaxed in the absence of the substrate and the metal~\cite{footnote_suspended},  and in the metal-deformed CNTs. After deformation, bandgap values are reduced by greater amounts in (3n+1,0) CNTs than  in (3n+2,0) CNTs, such that bandgap reductions fall in between 13\% for (17,0) and 31\% for (22,0) nanotubes. We find all collapsed CNTs to be metallic. We don't find drastic differences in the electronic structures as we slide atoms along the curvature direction, which is equivalent to rolling the tube along the surface obtained from a continuum model.

To explore the origins of the electronic structure modifications, we produced deformed CNTs structures within the continuum model by increasing metal surface energy $\gamma$ in Eq.~(\ref{energy_tot}) and setting van der Waals attraction between the sidewalls to zero, i.e. $U_{cc}=0$. The bandgaps gradually reduce with reducing the minimum radius of curvature $R_{min}$ and become zero at some critical value of $R_{min}\propto2.1-2.2$ \AA \ consistent with Ref.~\cite{NishidatePRB2008}, as shown in Fig.~\ref{fig:gap-rmin_open_and_close}. However, when we keep  $U_{cc}$ in the total energy minimization, we stabilize collapsed CNTs in which the bandgaps become zero at much larger  critical values of $R_{min}\propto2.4-2.5$ \AA, as shown in Fig.~\ref{fig:gap-rmin_open_and_close}. Moreover, we find that the bandgaps in (22,0) and (25,0) collapsed CNTs are much smaller than those in the open CNTs for the same values of $R_{min}$ due to the $\pi$-$\pi$ interactions in the former. Thus we conclude, that $\pi$-$\pi$ interactions between the orbitals on adjacent sidewalls separated  at the van der Waals distance in the collapsed CNTs produce much stronger electronic structure modifications than the curvature induced $\sigma$-$\pi$ interactions.

It is expected that changes in the electronic structure caused by the deformations would introduce contact resistance at the interface between the round and the deformed portions of a nanotube. To quantify the magnitude of resistance we
solve a scattering problem within a tight-binding formulation~\cite{groth2014kwant}. Following  Ref.~\cite{tomanek1988first},
we describe nanotube bandstructure  by a four-orbital tight-binding model to account for curvature-induced mixing of $\pi$ and $\sigma$ orbitals~\cite{blase1994hybridization}.
In the case of collapsed CNTs, an additional long-range interaction between carbons belonging to different layers is added~\cite{perebeinos2012phonon}.
The resulting band structure agrees fairly well with the DFT results, as shown in Fig.~\ref{fig:dft-and-TB}.

The contact resistances of the round-deformed interfaces in (19,0) and (25,0) nanotubes are shown in Fig.~\ref{fig:conductance}a and \ref{fig:conductance}b, correspondingly. As expected, in both cases the conductance is smaller than for a uniform tube, whether round or uniformly deformed, as shown by the dashed curves in Fig.~\ref{fig:conductance}. For the interface between the deformed and the round  portions of (19,0) nanotube, we see very little scattering due to  wavefunction mismatch, such that the contact conductance is around 95\% of the ideal conductance, at a typical metal-induced doping level of about 0.1-0.2 electrons per nm. However, a much stronger effect on the resistance is found for the round-collapsed contact interface. Lifting the double degeneracy alone in the fully collapsed (25,0) CNT results in a two times higher contact resistance as compared to the ideal resistance of $R_0=h/4e^2=6.5$ k$\Omega$. In addition, reflection at the interface causes an additional increase in the resistance.

\begin{figure}
\includegraphics[scale=0.32]{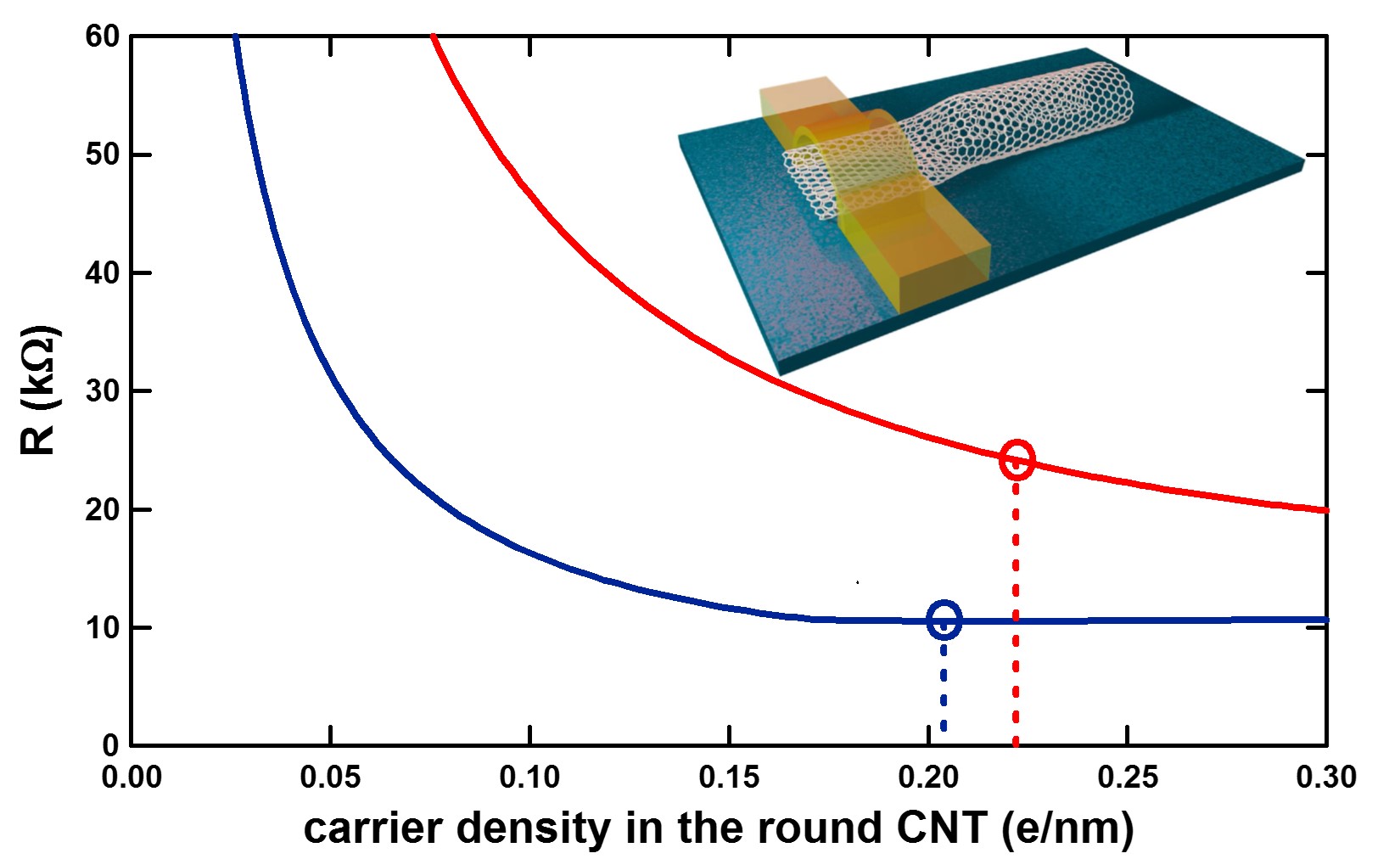}
\caption{Contact resistance in (19,0) (blue) and (25,0) (red) CNTs at $T=300$ K as a function of carrier density in the round portions of CNT. The Fermi level in CNT under the metal is assumed to be fixed, as explained in the text. The vertical dashed lines correspond to the estimated carrier densities in the on-state (see text). The inset shows a schematic of a CNT field-effect transistor contact~\cite{schematics}.}
\label{fig:resistance}
\end{figure}

In CNT field-effect transistor, schematic is shown in the inset of Fig.~\ref{fig:resistance}, the Fermi level can be modified by an applied gate voltage in the round portion of a CNT.
 Whereas under the metal, doping is determined by a CNT and metal work function difference, which for Pd contact we assume it to be  $\Delta W=0.35$ eV~\cite{Perebeinos2013CNTSchottky}.  The metal-CNT capacitance $C_M=2\pi\varepsilon_0/ln(1+2d_0/d)$ depends on CNT diameter $d$ and electrostatic distance $d_0=2.5$ \AA. We solve self-consistent equations for the charge density in CNT under the metal:
\begin{eqnarray}\label{rho_self}
  \rho &=& C_Me\phi= -\int^{\infty}_{E_{NP}} DOS(E-E_{NP}) f_0(E-E_F)dE
  \nonumber \\
  &+&\int_{-\infty}^{E_{NP}} DOS(E-E_{NP}) (1-f_0(E-E_F))dE
\end{eqnarray}
where charge neutrality point is $E_{NP}=\Delta W-e\phi$, $DOS(E)$ is the density of states from the DFT calculations, and $f_0$ is the Fermi-Dirac function. Doping determines the number of conduction modes $M$ at the Fermi energy $E_F$ and the maximum on-state conductance $G_{on}$~\cite{starvation_Guo,starvation_Fischetti,starvation_Wong} (in the absence of the tunneling current contribution~\cite{Perebeinos2013CNTSchottky}):
 \begin{eqnarray}\label{G_on}
  G_{on} &=&\frac{1}{R_{on}}=\frac{2e^2}{h}\int_{-\infty}^{\infty} T(E)M(E)\left(-\frac{\partial f_0}{\partial E}\right)dE
\end{eqnarray}
where $T(E)$ is  the transmission coefficient and the spin degeneracy is included. Using $DOS$ for a round CNT in Eq.~(\ref{rho_self}), the self consistent doping level would be $E_V-E_F=8$ meV, where $E_V$ is the top of the valence band,  and an ideal on-state resistance of $R_{on}=11.1$ k$\Omega$ follows from Eq.~(\ref{G_on}), using $T(E)=1$ and $T=300$ K. In the deformed CNT, the self consistent doping level is lower $E_V-E_F=20$ meV due to the smaller bandgap and the ideal on-state resistance is $R_{on}=9.5$ k$\Omega$. By calculating transmission coefficient $T(E)$ as a function of doping level in the round portion of the nanotube, which is controlled by the backgate voltage and which we model by introducing a rigid shift of the bandstructure in the round CNT with respect to that in CNT under the metal, we find resistance according to Eq.~(\ref{G_on}), shown in Fig.~\ref{fig:resistance}. We estimate a typical carrier density in the on-state using a wrap around SiO$_2$ gate with dielectric constant $\varepsilon=3.9$ and radius 20 nm, and the overdrive backgate voltage of 0.5 V.  As shown in Fig.~\ref{fig:resistance}, the on-state resistance of $R_{on}=10.5$ k$\Omega$ in (19,0) CNT is not very different from the ideal contact resistance from Eq.~(\ref{G_on}), using $T(E)=1$.

In the case of fully collapsed (25,0) CNT, the self-consistent Fermi energy of a round CNT under the metal would be  $E_V-E_F=36$ meV and the corresponding on-state resistance $R_{on}=8.1$ k$\Omega$. Using $DOS$ of the fully collapsed CNT, self consistent doping is found to be $E_V-E_F=208$ meV and the ideal contact resistance $6.5$ k$\Omega$ (should the double degeneracy be preserved). However, for the bandstructures corresponding to the collapsed CNT under the metal but round in the channel,  contact resistance calculations suggest a much higher contact resistance, as shown in Fig.~\ref{fig:resistance}. This is due to both degeneracy lifting and wavefunction mismatch. For a typical carrier density in the on-state,  we find contact resistance of $24$  k$\Omega$, three times higher compared to the ideal contact resistance from Eq.~(\ref{G_on}), using $T(E)=1$ and the doubly degenerate bands.

In conclusion, we identified the major effect of metal-induced nanotube deformations on the electronic structure and the electrical contact resistance at the interface between the deformed and the round portions of a nanotube. While the bandgap reduction in the deformed CNTs increases metal-induced doping of a nanotube, and thus reduces the resistance by increasing the number of the conduction channels,  wavefunction mismatch introduces an additional scattering at the contact, hence, partially compensating the effect of doping. Fully collapsed semiconducting nanotubes become metallic with a singly degenerate band at the Fermi level due to the broken axial symmetry.
The magnitude of the degeneracy lifting of few hundreds of meV is much larger than room temperature and that in the prior reports. Together with the wavefunction mismatch this leads to a much higher contact resistance as compared to the ideal contact. Our results may shed light to the observations of quantized conductance in large-diameter CNTs~\cite{BiercukPRL2005_CNT_quantized}  with steps quarter as large as usual and stimulate further experimental and theoretical work.


\begin{thebibliography}{45}
\bibitem [{for()}]{forceissue}%
  \BibitemOpen
  \href@noop {} {\bibinfo  {journal} { We find that forces along
  the CNT axis direction, can be very large up to 1.1 eV/\AA (or 70 GPa) and 2.6
  eV/\AA (or 160 GPa) in open (19,0) and collapsed (25,0) CNTs, correspondingly.
  This occurs in a very narrow region of the CNT circumference, where the
  radius of curvature is very small. This effect is not captured by the valence
  force model. By relaxing atoms in the z-direction along the nanotube axis,
  keeping the ($x, y$) coordinates the same in DFT, we confirmed that our main
  conclusions on the degeneracy lifting are not modified}\ }\BibitemShut
  {NoStop}%
\bibitem [{foo({\natexlab{b}})}]{footnote_suspended}%
  \BibitemOpen
\bibfield  {journal} {  }\href@noop {} {\bibfield  {journal} {\bibinfo
  {journal} {Bandstructure modifications caused by the subtrate-induced
  deformations alone are much smaller than those caused by the metal.}\ }
  ({\natexlab{b}})}\BibitemShut {NoStop}%
\bibitem [{sch()}]{schematics}%
  \BibitemOpen
  \href@noop {} {\bibinfo  {journal} { For the transport
  calculations we use an abrupt junction by setting up the corresponding tight
  binding matrix elements between the deformed and the round portions of the
  CNT}\ }\BibitemShut {NoStop}%
\end{thebibliography}
\end{document}